# Electrically packaged silicon-organic hybrid (SOH) I/Q-modulator for 64 GBd operation

HEINER ZWICKEL,[1] JUNED N. KEMAL,[1] CLEMENS KIENINGER,[1,2] YASAR KUTUVANTAVIDA,[1,2] JONAS RITTERSHOFER,[1] MATTHIAS LAUERMANN,[1,3] WOLFGANG FREUDE,[1] SEBASTIAN RANDEL,[1] AND CHRISTIAN KOOS[1,2,*]

[1]*Institute of Photonics and Quantum Electronics (IPQ), Karlsruhe Institute of Technology (KIT), Germany*
[2]*Institute of Microstructure Technology (IMT), Karlsruhe Institute of Technology (KIT), Germany*
[3]*Now with Vanguard Photonics GmbH, Karlsruhe, Germany*
*christian.koos@kit.edu

**Abstract:** Silicon-organic hybrid (SOH) electro-optic (EO) modulators combine small footprint with low operating voltage and hence low power dissipation, thus lending themselves to on-chip integration of large-scale device arrays. Here we demonstrate an electrical packaging concept that enables high-density radio-frequency (RF) interfaces between on-chip SOH devices and external circuits. The concept combines high-resolution $Al_2O_3$ printed-circuit boards with technically simple metal wire bonds and is amenable to packaging of device arrays with small on-chip bond pad pitches. In a set of experiments, we characterize the performance of the underlying RF building blocks and we demonstrate the viability of the overall concept by generation of high-speed optical communication signals. Achieving line rates (symbols rates) of 128 Gbit/s (64 GBd) using quadrature-phase-shift-keying (QPSK) modulation and of 160 Gbit/s (40 GBd) using 16-state quadrature-amplitude-modulation (16QAM), we believe that our demonstration represents an important step in bringing SOH modulators from proof-of-concept experiments to deployment in commercial environments.

## 1. Introduction

Efficient broadband electro-optic (EO) modulators are key components in microwave photonics and in optical communication systems. Over the last decade, the footprint of EO modulators has been radically reduced, moving from centimeter-long $LiNbO_3$-devices to millimeter- or sub-millimeter components that exploit high-index-contrast waveguides on semiconductor substrates [1–4]. In this context silicon photonics (SiP) has emerged as a particularly promising platform, exploiting advanced CMOS processes for high-yield mass production of modulators — either as standalone devices or as parts of more complex photonic integrated circuits (PIC) [5–8]. To improve the performance of SiP modulators, silicon nanowire waveguides can be combined with highly efficient organic EO materials that have been engineered on a molecular level [9,10]. This so-called silicon-organic hybrid (SOH) approach allows to reduce the voltage-length product of EO modulators to 0.32 Vmm [11] — more than an order of magnitude below that of conventional depletion-type SiP modulators. SOH integration allows to realize high-speed modulators with sub-millimeter device lengths. The modulators offer line rates up to 120 Gbit/s for intensity modulation and direct detection (IM/DD) [12], and up to 400 Gbit/s for coherent 16QAM signaling [13].

However, while reducing the on-chip footprint of EO modulators increases integration density as well as modulation speed, electrical packaging of the devices becomes increasingly difficult [14]. One of the main problems is the vast size mismatch between the ultra-fine features that can be realized by advanced deep-UV lithography on the silicon chip and the much larger dimensions of radio frequency (RF) traces that are commonly available on printed circuit boards (PCB) or interposers. This often results in a mismatch of bond pad pitch, which impedes high-density packaging of large-scale device arrays and must be overcome by excessively long metal wire bonds with limited performance in terms of bandwidth and signal fidelity [15]. In the field of microelectronics, flip-chip bonding is often used for broadband connections between interposers and densely integrated on-chip circuits [16]. This approach has been recently transferred to packaging of advanced SiP transceivers [17–19], enabling transmission of optical signals with data rates of up to 136 Gbit/s using 16QAM signaling at a symbol rate of 34 GBd [19]. However, flip-chip bonding of PIC blocks access to the chip surface and thus complicates optical packaging, often leaving edge coupling to actively aligned optical fibers as the only option [20]. Moreover, flip-chip bonding of photonic devices often requires dedicated processes that need to be adapted to specific optical components in terms of thermal budget or with respect to the deployed flux chemicals [15,21]. In contrast to flip-chip bonding, accessing high-density PIC by metallic wire bonds leaves the top surface of the chip free and thus offers high flexibility with respect to optical packaging [22–30], exploiting, e.g., highly efficient SiP grating couplers [31], photonic wire bonds [32], or facet-attached micro-lenses [33]. However, published signaling demonstrations using SiP modulators accessed by metal wire bonds have so far been limited to symbol rates of 28 GBd. Moreover, none of the aforementioned approaches have so far been applied to SOH devices.

In this paper, we demonstrate an electrically packaged SOH EO modulator that exploits simple metal wirebonds in combination with a high-resolution ceramic PCB to establish broadband connections from standard coaxial connectors to sub-mm on-chip devices. Our approach allows to adapt the bond-pad pitch on the PCB to that of the PIC, thereby permitting connections through short parallel metal wire bonds that are amenable to broadband packaging of large-scale device arrays. In a set of experiments, we expand upon our earlier work [34] and demonstrate line rates (symbols rates) of 128 Gbit/s (64 GBd) using quadrature-phase-shift-keying (QPSK) modulation, and line rates (symbol rates) of 160 Gbit/s (40 GBd) using 16-state quadrature-amplitude-modulation (16QAM). This is among the highest symbol and line rates demonstrated so far with electrically packaged SiP modulators. In addition, our devices feature a voltage-length product of only $U_\pi L = 0.9$ Vmm, which is well below that of packaged SiP devices that were previously demonstrated. When combined with recent progress in improving thermal stability of organic EO materials [35] and of SOH modulators [36] as well as with novel technologies for hybrid photonic integration [37], we believe that our demonstration represents an important step in bringing SOH modulators from proof-of-concept experiments to deployment in laboratory and commercial environments.

## 2. Concept, fabrication, and assembly

This section introduces the concept of silicon-organic hybrid (SOH) modulators, reports on the fabrication and characterization of the high-frequency ceramic printed circuit board (PCB), and shows the assembly of the electrically packaged modulator.

### 2.1 SOH Mach-Zehnder modulator

The concept of a strip-loaded SOH Mach-Zehnder modulator (MZM) [9,38,39] is depicted in Fig. 1. The basic optical waveguide structure is fabricated on a standard 220 nm silicon-on-insulator (SOI) wafer using 248 nm deep-UV lithography. The MZM consists of two parallel 0.6 mm long phase shifter sections. Each phase shifter comprises a silicon slot waveguide filled with an electro-optic (EO) organic material. The slot has a width of 120 nm. The 240 nm wide Si rails of the slot waveguide are connected to the electrodes with thin, doped Sis labs and Al vias. The device is realized as traveling-wave modulator where the optical mode in the slot waveguide co-propagates with the electric (modulating) mode. Both phase shifter sections share the central signal electrode which, together with the outer ground electrodes, form a coplanar ground-signal-ground transmission line. Aluminum vias connect the electrodes to bond pads of the top metal layer at both ends of the transmission line. The electrical RF voltage drops mainly across the narrow slot where the light is highly confined, thus leading to a high overlap of the modulating electric field and the optical mode and a high modulation efficiency. As an EO material, we use the guest-host material SEO250, which is locally deposited on the slot waveguides by a high-resolution dispensing technique. To activate the macroscopic $\chi^{(2)}$-nonlinearity, the microscopic molecular dipole moments of chromophores in the organic EO material need to be aligned in a one-time poling process. To this end, a poling voltage $U_{pol}$ is applied across the (floating) ground electrodes at an elevated temperature for aligning the EO chromophores. After cooling the device down, the poling voltage is removed, and the molecular orientation is frozen as indicated by the green arrows in Fig. 1(e-3). The modulating field $E_{mod}$ induced by the RF drive voltage is oriented parallel to the chromophore alignment in one arm of the MZM and antiparallel in the other arm, which leads to chirp-free push-pull operation. The PIC used in this work contains an array of MZM that are configured as nested pairs to serve as IQ-modulators. Alternatively, the unused ports of the MZM can be accessed directly via attached grating couplers, which allows to use the devices individually for amplitude or intensity modulation.

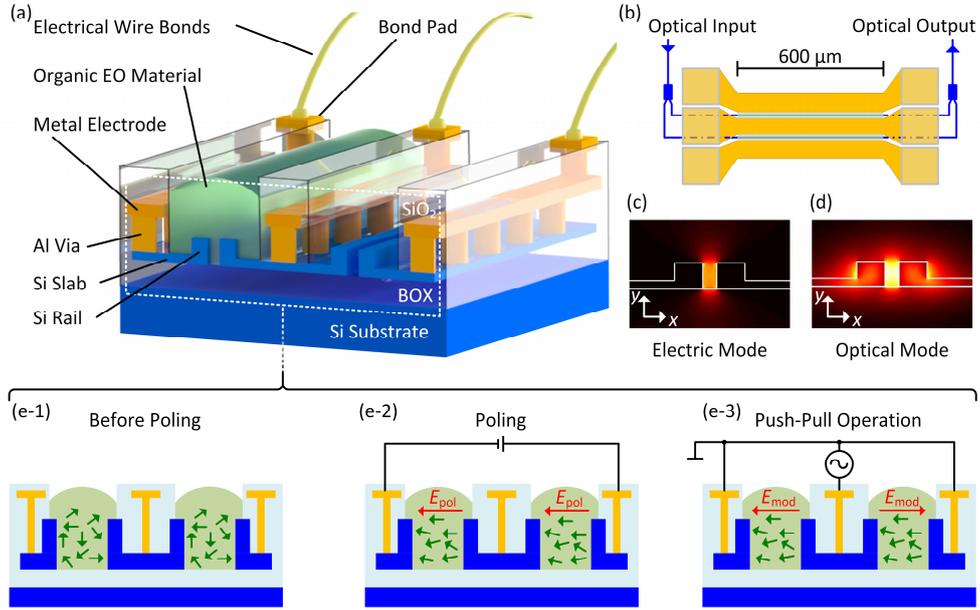

Fig. 1. Device concept and operation of an SOH Mach-Zehnder modulator (MZM). **(a)** Artist's view of the MZM: The organic EO cladding (green) is shown in one arm only. Aluminum vias connect the doped thin Si slabs to the traveling-wave electrodes and to the bond pads on the top metal layer. **(b)** Schematic of traveling-wave MZM with electrodes that form a ground-signal-ground coplanar transmission line having bond pads at both ends. **(c, d)** Colour plot of the dominant *x*-component of the RF electric field and the optical electric field. The strong overlap leads to efficient modulation. **(e-1) to (e-3)**: Poling of the modulator. **(e-1)** Prior to poling, the molecular dipoles (green arrows) in the organic EO material are randomly oriented such that no macroscopic Pockels effect can be observed. **(e-2)** At elevated temperatures, a poling voltage induces a poling field $E_{pol}$ which aligns the molecular dipoles predominantly along the electric field. This molecular orientation is frozen when the material is cooled down to room temperature. **(e-3)** In push-pull operation, the modulating RF voltage is applied to the center signal electrode thereby leading to a modulating field $E_{mod}$ that is oriented parallel to the poling direction in one arm and anti-parallel in the other one. This allows for pure amplitude or intensity modulation without residual phase modulation (chirp-free) [12].

### 2.2 High-frequency PCB

The PCB traces are realized as 50 Ω coplanar ground-signal-ground (GSG) transmission lines (TL). To facilitate electrical board-to-chip connections with short wire bonds and to enable the integration of densely spaced modulator arrays, the spacing of the board-level RF traces must be matched to the pitch of the bond pads on the silicon chip, which is chosen to be 100 μm to save chip area while enabling reliable placement of metal wire bonds. At the same time, the board-level RF traces must allow for easy interfacing to discrete devices such as coaxial connectors or surface mount devices (SMD), which feature millimeter-size contact spacing. To this end, we use tapered sections of RF traces with impedance-matched transitions between RF traces of different dimensions. Fabrication of the underlying PCB requires structuring techniques that combine micrometer resolution with the ability to process boards with centimeter-scale overall dimensions. To this end, we have developed a direct-laser write (DLW) fabrication process which allows for flexible mask-less fabrication at low cost. The process starts from square $100 \times 100$ mm$^2$ Al$_2$O$_3$ ceramic substrates which are covered by a 3 μm thick electroplated Au layer. The substrates are coated with a positive-tone photoresist (*AZ1500*) and patterned with a Nd:YAG laser. After development of the exposed structures, the Au layer is wet-etched using a KI solution, and the resist mask is removed.

The performance of the RF building blocks is characterized using dedicated test structures. Figure 2(a) shows the TL near the chip edge, Fig. 2(b) displays the taper section, and Fig. 2(c) illustrates the TL near a coaxial connector. For the characterization, the TL as used near the chip edge is contacted on both sides using GSG microwave probes, and the *S*-parameters are measured using a vector network analyzer (VNA). The reference planes are moved to the probe tips by using a short/open/load/through (SOLT) calibration routine with a proper calibration substrate. Using the method described in [40], we extract the characteristic impedance $Z_0$ and the power loss coefficient *α* from the measured *S*-parameters, Fig. 2(d).

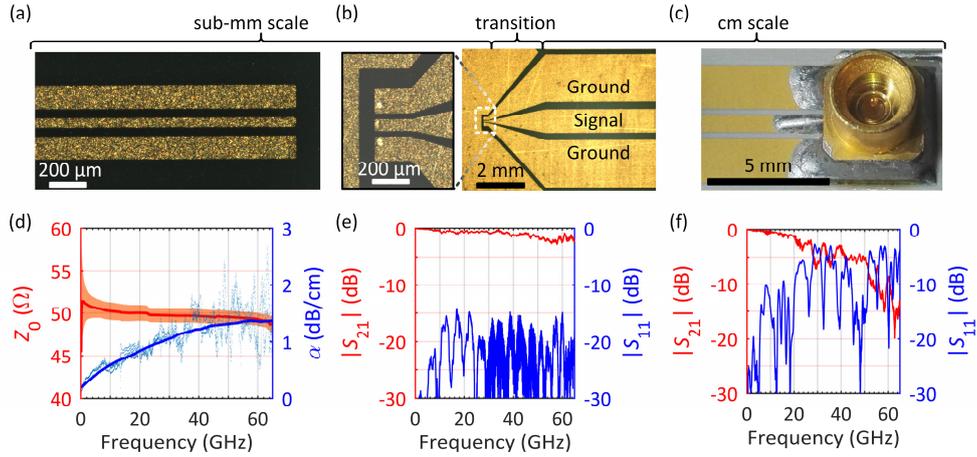

Fig. 2. Characterization of the RF building blocks. (a) Micrograph of a coplanar TL with dimensions matching the bond pads on the Si chip. The TL has been fabricated on gold-plated ceramic substrates using direct laser writing and wet etching. (b) Micrograph of a tapered section of the TL. The left-hand side corresponds to the TL shown in (a), the right-hand side matches the TL shown in (c). (c) Test structure to characterize the transition from a surface-mounted coaxial connectors to a TL. (d) Characteristic impedance $Z_0$ (dark red: mean value from a set of 10 measurements, light red: corresponding standard deviation) and power propagation loss *α* (light blue dots: measurement data; blue line: moving average as a guide to the eye) extracted from *S*-parameters of TL with the small cross-section shown in (a). The characteristic impedance $Z_0$ is well matched to 50 Ω, and the losses stay below 2 dB/cm. (e) *S*-parameters of a test structure comprising a 9 mm long TL with large cross section as shown in (c), terminated by a pair of tapers as shown in (b). The measurements exhibit the high transmission and reflection quality of the tapers. (f) *S*-parameters obtained from a test structure comprising a 7.8 mm long TL terminated by a pair of coaxial SMP-M connectors as shown in (c). The measurement also comprised the influence of two semi-flexible cable assemblies converting from the 1.85 mm VNA connectors to the SMP-M interface.

The characteristic impedance $Z_0$ is well matched to the system impedance of 50 Ω over the entire measurement range. The power loss coefficient $α$, presumably dominated by conductor loss, increases with frequency due to the skin effect, but stays below 2 dB/cm, which compares well to the losses achieved on state-of-the-art calibration substrates [41]. The characteristic impedance and the power loss coefficient shown in Fig. 2(d) were extracted from a set of 10 different TL, all of which yield comparable results.

The transition between TL of different cross sections is realized by an impedance-matched linear taper. Figure 2(e) shows the measured *S*-parameters of the corresponding test structure comprising two tapers at both sides of a 9 mm long TL section as in Fig. 2(b). The 3-dB bandwidth extracted from $S_{21}$ is larger than 60 GHz, and the reflection as expressed by $S_{11}$ stays below −15 dB.

To feed the signal from coaxial cables to the PCB, surface-mounted coaxial connectors are soldered to appropriate pads connected to the centimeter-scale transmission line. Measured *S*-parameters of a structure comprising two such connectors (surface-mounted type SMP-M) at the ends of a 7.8 mm long TL indicate good performance up to 20 GHz. However, the increase of the power reflections limits the performance in the 25…40 GHz range, see Fig. 2(f). The reduction in transmitted power beyond 25 GHz is attributed to losses in the coaxial connectors and the transition to the on-board pads. Note that our measurements with SMP-M connectors include the influence of two semi-flexible cable assemblies converting from 1.85 mm VNA connectors to the SMP-M interface, which we could not isolate because we lack an SMP-M calibration kit. These assemblies are specified to have a 3dB bandwidth > 65 GHz such that we do not expect any significant distortions of our measurements.

*Assembly*

Figure 3 shows the module assembly. The SiP chip is glued on a metal sub-mount using a conductive epoxy adhesive, Fig. 3. After poling the modulators, two identical ceramic PCB are placed on both sides of the Si chip and are glued to the same metallic sub-mount (see Fig. 3(a) and Fig. 3(c)). By precise machining of the sub-mount, the surfaces of the Si chip and the PCB are brought to approximately the same height to allow for short wire bonds. To avoid excitation of unwanted substrate modes, the sub-mount features a void below the RF traces, see Fig. 3(a). The traveling-wave electrodes of the modulators are connected to the PCB traces at both ends using Au wires attached by wedge-to-wedge bonding. Wedge bonding was chosen because the wire lengths can be kept shorter than with ball bonding. A short bondwire connection is desirable for high-frequency operation. Surface-mounted GPPO-compatible SMP-M receptacles and coaxial cables are used to connect to the signal source at one end of the transmission line and to a terminating resistor at the other end. The module contains two MZM that can be used either individually for amplitude or intensity modulation, or, by using a different optical input/output, as two nested MZM which form an I/Q-modulator. Figure 3(c) shows an image of the module, and Fig. 3(d) displays a micrograph of the bonded chip. Light is coupled to and received from the chip using on-chip grating couplers.

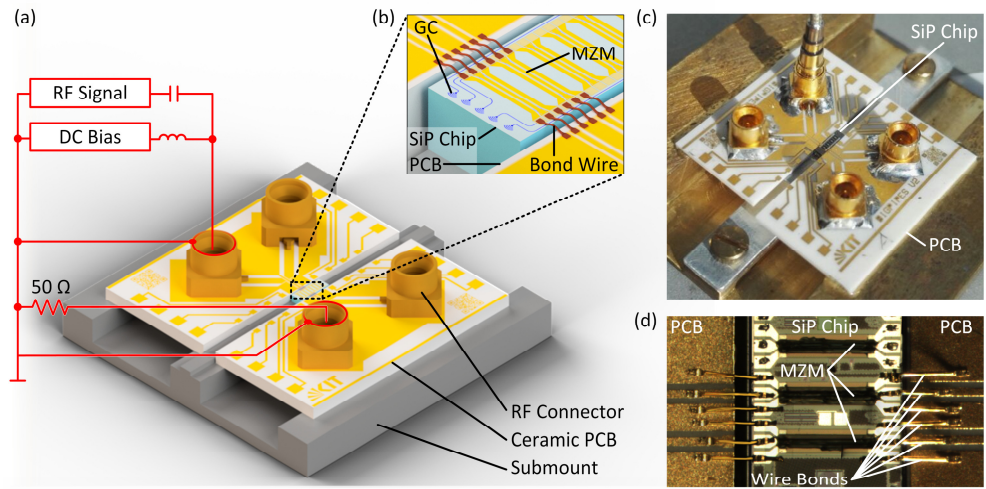

Fig. 3. Electrical packaging. (a) Artist's view of the module connecting two silicon-organic hybrid (SOH) Mach-Zehnder modulators (MZM) to surface-mounted SMP-M connectors on a ceramic printed circuit board (PCB). The external electrical circuit used for experiments is depicted schematically in red for one of the MZM (b) Two MZM, connected by bond wires. (c) Image of the complete module. (d) Micrograph of the SiP chip. Wire bonds are used for connecting the chip to the traces on the two ceramic PCB.

## 3. Optical, DC, and small-signal characterization of module

To measure the optical transmission spectrum of the electrically packaged MZM, we use a tunable laser. The static extinction ratio of the modulator exceeds 30 dB. The fiber-to-fiber insertion loss is 11.6 dB. This includes overall fiber-chip coupling losses of 8.9 dB caused by the two grating couplers, as well as an on-chip device loss of 1.3 dB, caused by passive components, namely two MMI splitters (0.37 dB each), input/output access waveguides (0.2 dB in total), and mode converters at the strip-slot waveguide interface (0.4 dB in total). The remaining 1.4 dB are attributed to slot waveguide loss in the phase shifter sections, which leads to an estimated propagation loss of $a$ = 2.3 dB/mm. The π-voltage $U_\pi$ of the MZM is measured by applying a low-frequency (< 1 kHz) electrical signal, and by recording the applied voltage and the transmitted optical power using an oscilloscope. The π-voltage $U_\pi$ corresponds to the voltage that needs to be applied to the center electrode of the MZM in the low-frequency limit to induce a relative phase shift of π between the two arms of the device. We find $U_\pi$ = 1.5 V, leading to a modulation efficiency of the 0.6 mm long MZM of $U_\pi L$ = 0.9 Vmm and a π-voltage-loss product $aU_\pi L$ = 2.1 dBV — the best value so far achieved in SOH modulators using an EO material with a high glass transition temperature $T_g$. Note that this π-voltage-loss product is significantly smaller than the best value achieved in high-speed depletion-type SiP modulators, which amounts to 12 dBV [42]. The low $U_\pi L$-product of the SOH modulator results from the high EO activity of the organic cladding material in combination with the slot-waveguide geometry of the phase shifter [9], which leads to a good overlap of the optical mode and the modulating RF field. To improve the EO modulation bandwidth of the MZM, the conductivity of the doped silicon slabs can be increased by a gate voltage applied between the ground electrode and the Si substrate [43]. This leads to an accumulation layer of electrons at the interface of the doped Si slabs and the buried oxide, thereby increasing the slab conductivity. In the experiment, we use a gate field strength of 50 V/μm, which increases the loss of the slot waveguide only moderately by 0.2 dB. This setting was used for the dynamic analysis shown below, and for the data generation (Section 4). Note that the gate voltage can be reduced by using a gate electrode [44] or even omitted by using optimized doping profiles. The dynamic behavior of the MZM is characterized for the small-signal case using a VNA. First, the electrical *S*-parameters of the unpackaged MZM are measured using microwave probes at both ends of the modulator's coplanar TL. The result is shown in Fig. 4(a) along with the *S*-parameters of the electrically packaged modulator, which were measured by connecting the VNA ports directly to the SMP-M connectors on the ceramic PCB. The package has a significant influence on the electrical transmission spectrum at frequencies larger than 20 GHz. This is in agreement with the finding from Fig. 2(f), where a drop of the transmission in the transition to the surface-mounted connectors limits the performance beyond 20 GHz. In the low-frequency range, no significant impact on the EO performance is expected. This is confirmed by a second experiment, in which the electro-optical-electrical (EOE) bandwidth is measured using a VNA and a calibrated photodiode. Figure 4(b) shows the EOE responses of the electrically packaged and unpackaged MZM.

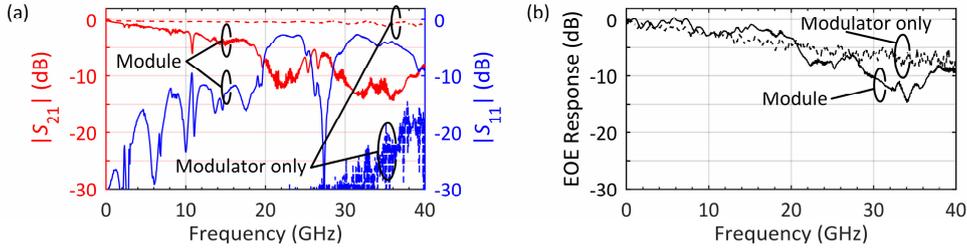

Fig. 4. Small-signal characterization of the MZM module. (a) Electrical *S*-parameters of the unpackaged modulator measured with microwave probes (dashed line), and of the electrically packaged modulator measured by connecting the VNA ports directly to the SMP-M connectors on the ceramic PCB (solid lines). (b) Electro-optical-electrical (EOE) bandwidth of the unpackaged modulator (dashed line), and of the electrically packaged modulator (solid line). Both *S*-parameters traces are normalized to the respective value measured at 40 MHz. At frequencies beyond 20 GHz, a slight roll-off is observed for the packaged device with respect to the unpacked device. The electro-optic 3-dB (6-dB) bandwidth for the electrically packaged modulator is 21 GHz (31 GHz).

As expected, the EOE responses are comparable up to a frequency of 20 GHz, with a rather modest frequency roll-off beyond that frequency. There is even a slight improvement of the frequency roll-off in the low-frequency range for the packaged MZM with respect to the unpackaged device. This observation is attributed to resonant peaking caused by imperfect impedance matching. The electro-optic 3-dB (6-dB) bandwidth for the electrically packaged modulator is 21 GHz (31 GHz). At frequencies between 20 GHz and 30 GHz, only a slight roll-off is observed for the packaged device with respect to the unpacked device. Beyond 30 GHz, the roll-off of the packaged device is more pronounced, thus limiting the achievable symbol rates in comparison the earlier transmission demonstrations that were performed with unpackaged devices contacted by high-frequency microwave probes.

Note that the response of the Pockels-type EO material in the modulator is ultra-fast and would in principle allow for THz modulation bandwidths. In real SOH devices, however, the speed is limited by the capacitance of the slot waveguide that needs to be charged and discharged via the resistive doped silicon slabs [9]. This effect can be mitigated by an improved doping profile for the silicon slabs, which reduces the resistivity and the associated RC time constant and therefore improves the bandwidth of the modulator without an undue increase of the optical loss. In addition, we may improve the bandwidth of the packaged module by using a more elaborate design of the critical transition between the coaxial connector and the coplanar transmission line on the PCB. Note that, besides the EOE bandwidth itself, the characteristic of the transfer function beyond the corner frequency is also important. As an example, a modulator with a bandwidth of 25 GHz but with a moderate roll-off can easily support a 100 Gbit/s OOK modulation [45].

### 4. Intensity-modulated and coherent signaling

Using a single MZM, intensity-modulated signals are generated in a first set of experiments. Four channels of a pulse pattern generator (PPG) are serialized in an electrical 4 × 1 multiplexer (MUX) to generate a pseudo random binary sequence (PRBS) with a length of $2^{31}-1$. The output of the MUX delivers a voltage swing of 2 $V_{pp}$ and is directly connected to the input of the module. Light is coupled to the device using grating couplers. After the modulator, the optical signal is amplified by an erbium-doped amplifier (EDFA). Out-of-band amplified spontaneous emission (ASE) noise is removed using an optical band-pass filter. The signal is detected using a photodiode connected to an equivalent-time sampling oscilloscope (*Agilent 86100C*), exploiting the repetitive nature of the test pattern for broadband electrical acquisition. Recorded eye diagrams at 20 Gbit/s and 40 Gbit/s are shown in Fig. 5(a) and Fig. 5(b), respectively. At 20 Gbit/s (40 Gbit/s) the measured *Q*-factor amounts to 9.8 (4.1).

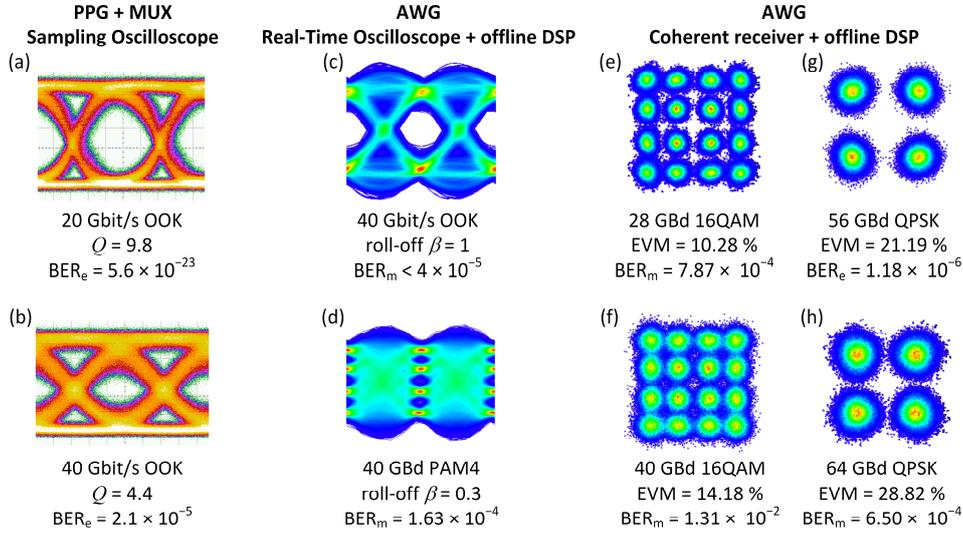

Fig. 5. Optical signaling using various modulation formats. (a) Eye diagrams for 20 Gbit/s OOK and (b) 40 Gbit/s OOK without pre-compensation or post-processing using a pulse-pattern generator and an equivalent-time sampling oscilloscope. (c) Eye diagrams for 40 Gbit/s OOK with an adaptive equalizer applied on the recorded data using a real-time sampling oscilloscope and offline DSP. Pulses with a raised-cosine spectrum and a roll-off factor of $\beta = 1$ were used. (d) Eye diagram for 40 GBd (80 Gbit/s) PAM4 using pulse shaping with $\beta = 0.3$ and offline DSP. (e-h) Constellation diagrams for (e) 28 GBd 16QAM, (f) 40 GBd 16QAM, (g) 56 GBd QPSK and (h) 64 GBd QPSK.

Neglecting inter-symbol interference (ISI) and assuming additive Gaussian noise as the only signal distortion, a bit error ratio (BER) of $BER_e = 5.6 \times 10^{-23}$ at 20 Gbit/s is estimated [46], and a $BER_e = 2.1 \times 10^{-5}$ at 40 Gbit/s. Note that the underlying assumptions might not hold exactly, especially at 40 Gbit/s where ISI might start to play a role. Still, without claiming exact quantitative accuracy of the estimated $BER_e$ values, we are confident that the true BER falls well below the forward error correction (FEC) threshold of $4.40 \times 10^{-3}$ for less than 7% overhead [47].

In a second set of experiments, an arbitrary waveform generator (AWG, *Keysight M8196A*) replaces the PPG, and a real-time sampling oscilloscope (*Keysight DSO-X 93204A*) replaces the equivalent-time sampling oscilloscope of the previous experiment. In a first step, we generate intensity-modulated two- and multilevel signals, record the received data, and apply digital signal processing (DSP) for equalization and analysis of the received signal. The AWG generates a PRBS with a length of $2^{15}-1$. The resulting pulses are shaped to have root-raised cosine spectrum. Figure 5(c) shows the eye diagram of a 40 Gbit/s OOK signal, generated with a roll-off factor $\beta = 1$, after applying an adaptive equalizer to flatten out the end-to-end frequency response of the system using an oversampling factor of ten. Here, the SOH MZM was driven with a peak-to-peak voltage swing of 1.4 $V_{pp}$ at the input coaxial connector of the module. No errors were found in our 62 μs long recording ($2.48 \times 10^6$ measured bit), which confirms that the BER is below $4 \times 10^{-5}$. For OOK at a line rate of 56 Gbit/s and using $\beta = 0.1$, a $BER_m = 1.96 \times 10^{-5}$ is measured (not displayed in the figure). Figure 5(d) shows the eye diagram for a four-level pulse amplitude (PAM4) modulated signal, generated using $\beta = 0.3$, at a symbol rate of 40 GBd (corresponding to a line rate of 80 Gbit/s). The BER is measured to be $BER_m = 1.63 \times 10^{-4}$, well below the 7% overhead FEC threshold.

The IM/DD data generation is complemented by the generation of quadrature-amplitude-modulated (QAM) signals, driving the electrically packaged SOH IQ modulator by a two-channel AWG and detecting the signal by a coherent receiver. Figure 5(e) shows the constellation diagrams for a 28 GBd 16QAM signal. The BER is measured to be $BER_m = 7.78 \times 10^{-4}$. For 40 GBd, see Fig. 5(f), the BER increases to $1.31 \times 10^{-2}$ but stays below the threshold of soft-decision FEC codes with a 20 % overhead [47]. This results in a line rate of 160 Gbit/s and a post-FEC net data rate of 133 Gbit/s on a single polarization. Using QPSK signals, symbol rates of 56 Gbd and 64 GBd were generated with a BER below the threshold for a hard-decision FEC with 7% overhead, see Fig. 5(g) and Fig. 5(h). Note that at a symbol rate of 64 GBd, we used a PRBS of length $2^{11}-1$ due to hardware limitations of our AWG. In all data generation experiments, a gate field of 50 V/μm has been applied to the modulator substrate [44].

These results compare well with previously published signaling demonstrations using SiP modulators accessed by metal wire bonds, for which 16QAM symbol rates of 28 GBd and line rates of 112 Gbit/s were demonstrated for a single polarization. Relying on the published state-of-the art, we believe that our demonstrations of 64 GBd QPSK and of 40 GBd 16QAM correspond to the highest symbol rate and the highest line rate achieved so far with a wire-bonded SiP modulators [29] and that the approach can well compete with more complex packaging concept based on flip-chip bonding [19]. In comparison to earlier demonstrations of packaged pn-depletion-type MZM, our SOH device features a significantly lower voltage-length product of only $U_\pi L = 0.9$ Vmm, well below that of best-in-class depletion-type SiP modulators with $U_\pi L = 6$ Vmm [42].

## 5. Summary

We demonstrate for the first time that electrical packaging of high-speed SOH modulators is possible without significantly degrading the performance of the devices. To this end, we developed a dedicated packaging concept that utilizes technically simple metal wire bonding for packaging of highly compact SOH devices. The concept relies on high-resolution ceramic PCB that allow to adapt the interposer bond-pad pitch to that of the PIC, thereby permitting connections through short parallel metal wire bonds that are amenable to broadband packaging of densely integrated device arrays. The modulators feature a π-voltage-length product of $U_\pi L = 0.9$ Vmm and a π-voltage-loss product of $aU_\pi L = 2.1$ dBV, thereby outperforming conventional depletion-type pn-modulators. In our experiments, we demonstrate generation of intensity modulated signals with line rates of up to 80 Gbit/s as well as of QPSK and 16QAM signals with symbol rates up to 64 GBd and lines rates up to 160 Gbit/s, respectively. In combination with recent progress regarding EO material stability [36] and novel technologies for hybrid photonic integration [37], our concept builds upon the intrinsic efficiency and performance advantages demonstrated for die-level SOH devices and allows to exploit them on the level of electrically packaged communication modules.

## Funding

Deutsche Forschungsgemeinschaft (DFG) project HIPES (383043731); European Research Council (ERC Consolidator Grant "TeraSHAPE" (773248); ERC Proof-of-Concept Grant 'SCOOTER', (755380)); EU-FP7 projects PhoxTroT (318240) and BigPipes (619591); Alfried Krupp von Bohlen und Halbach Foundation; Helmholtz International Research School for Teratronics (HIRST); Karlsruhe School of Optics and Photonics (KSOP); and Karlsruhe Nano-Micro Facility (KNMF).

## Acknowledgment

We thank Jingdong Luo and Alex K.-Y. Jen from Soluxra for providing the organic EO material. We thank Oswald Speck, Florian Rupp, Marco Hummel und Andreas Lipp for support during packaging of the modulator.

Portions of this work were presented at the Conference on Lasers and Electro-Optics (CLEO), 2018 in San José, paper SM3B.1. [34].

## References

1. S. Lange, S. Wolf, J. Lutz, L. Altenhain, R. Schmid, R. Kaiser, M. Schell, C. Koos, and S. Randel, "100 GBd intensity modulation and direct detection with an InP-based monolithic DFB laser Mach–Zehnder modulator," J. Lightwave Technol. **36**(1), 97–102 (2018).
2. P. C. Schindler, D. Korn, C. Stamatiadis, M. F. OKeefe, L. Stampoulidis, R. Schmogrow, P. Zakynthinos, R. Palmer, N. Cameron, Y. Zhou, R. G. Walker, E. Kehayas, S. Ben-Ezra, I. Tomkos, L. Zimmermann, K. Petermann, W. Freude, C. Koos, and J. Leuthold, "Monolithic GaAs electro-optic IQ modulator demonstrated at 150 Gbit/s with 64QAM," J. Lightwave Technol. **32**(4), 760–765 (2014).
3. G. T. Reed, G. Mashanovich, F. Y. Gardes, and D. J. Thomson, "Silicon optical modulators," Nat. Photonics **4**(8), 518–526 (2010).
4. A. Abbasi, J. Verbist, L. A. Shiramin, M. Verplaetse, T. De Keulenaer, R. Vaernewyck, R. Pierco, A. Vyncke, X. Yin, G. Torfs, G. Morthier, J. Bauwelinck, and G. Roelkens, "100-Gb/s electro-absorptive duobinary modulation of an InP-on-Si DFB laser," IEEE Photonics Technol. Lett. **30**(12), 1095–1098 (2018).
5. D. Inniss and R. Rubenstein, *Silicon Photonics: Fueling the Next Information Revolution,* (Morgan Kaufmann, 2016).
6. A. H. Atabaki, S. Moazeni, F. Pavanello, H. Gevorgyan, J. Notaros, L. Alloatti, M. T. Wade, C. Sun, S. A. Kruger, H. Meng, K. Al Qubaisi, I. Wang, B. Zhang, A. Khilo, C. V. Baiocco, M. A. Popović, V. M. Stojanović, and R. J. Ram, "Integrating photonics with silicon nanoelectronics for the next generation of systems on a chip," Nature **556**(7701), 349–354 (2018).
7. T. Baehr-Jones, T. Pinguet, P. Lo Guo-Qiang, S. Danziger, D. Prather, and M. Hochberg, "Myths and rumours of silicon photonics," Nat. Photonics **6**(4), 206–208 (2012).
8. A. Rickman, "The commercialization of silicon photonics," Nat. Photonics **8**(8), 579–582 (2014).
9. C. Koos, J. Leuthold, W. Freude, M. Kohl, L. Dalton, W. Bogaerts, A. L. Giesecke, M. Lauermann, A. Melikyan, S. Koeber, S. Wolf, C. Weimann, S. Muehlbrandt, K. Koehnle, J. Pfeifle, W. Hartmann, Y. Kutuvantavida, S. Ummethala, R. Palmer, D. Korn, L. Alloatti, P. C. Schindler, D. L. Elder, T. Wahlbrink, and J. Bolten, "Silicon-organic hybrid (SOH) and plasmonic-organic hybrid (POH) integration," J. Lightwave Technol. **34**(2), 256–268 (2016).
10. W. Heni, Y. Kutuvantavida, C. Haffner, H. Zwickel, C. Kieninger, S. Wolf, M. Lauermann, Y. Fedoryshyn, A. F. Tillack, L. E. Johnson, D. L. Elder, B. H. Robinson, W. Freude, C. Koos, J. Leuthold, and L. R. Dalton, "Silicon-organic and plasmonic-organic hybrid photonics," ACS Photonics **4**(7), 1576–1590 (2017).
11. C. Kieninger, Y. Kutuvantavida, D. L. Elder, S. Wolf, H. Zwickel, M. Blaicher, J. N. Kemal, M. Lauermann, S. Randel, W. Freude, L. R. Dalton, and C. Koos, "Ultra-high electro-optic activity demonstrated in a silicon-organic hybrid modulator," Optica **5**(6), 739–748 (2018).
12. H. Zwickel, S. Wolf, C. Kieninger, Y. Kutuvantavida, M. Lauermann, T. de Keulenaer, A. Vyncke, R. Vaernewyck, J. Luo, A. K.-Y. Jen, W. Freude, J. Bauwelinck, S. Randel, and C. Koos, "Silicon-organic hybrid (SOH) modulators for intensity-modulation / direct-detection links with line rates of up to 120 Gbit/s," Opt. Express **25**(20), 23784–23800 (2017).
13. S. Wolf, H. Zwickel, C. Kieninger, M. Lauermann, W. Hartmann, Y. Kutuvantavida, W. Freude, S. Randel, and C. Koos, "Coherent modulation up to 100 GBd 16QAM using silicon-organic hybrid (SOH) devices," Opt. Express **26**(1), 220–232 (2018).
14. T. Tekin, "Review of packaging of optoelectronic, photonic, and MEMS components," IEEE J. Sel. Top. Quantum Electron. **17**(3), 704–719 (2011).
15. L. Carroll, J.-S. Lee, C. Scarcella, K. Gradkowski, M. Duperron, H. Lu, Y. Zhao, C. Eason, P. Morrissey, M. Rensing, S. Collins, H. Hwang, and P. O'Brien, "Photonic packaging: transforming silicon photonic integrated circuits into photonic devices," Appl. Sci. (Basel) **6**(12), 426 (2016).
16. P. Elenius and L. Levine, "Comparing flip-chip and wire-bond interconnection technologies," Chip Scale Review **4**, 81 (2000).
17. D.-W. Kim, A. L. E. Jin, M. K. Raja, V. V. Kulkarni, L. C. Wai, J. L. T. Yang, and P. L. G. Qiang, "Compactly packaged high-speed optical transceiver using silicon photonics ICs on ceramic submount," in *2016 IEEE 66th Electronic Components and Technology Conference (ECTC)* (IEEE, 2016), pp. 1075–1080.
18. C. Doerr, J. Heanue, L. Chen, R. Aroca, S. Azemati, G. Ali, G. McBrien, L. Chen, B. Guan, H. Zhang, X. Zhang, T. Nielsen, H. Mezghani, M. Mihnev, C. Yung, and M. Xu, "Silicon photonics coherent transceiver in a ball-grid array package," in *Optical Fiber Communication Conference (OFC)* (OSA, 2017), paper Th5D.5.
19. A. Novack, M. Streshinsky, T. Huynh, T. Galfsky, H. Guan, Y. Liu, Y. Ma, R. Shi, A. Horth, Y. Chen, A. Hanjani, J. Roman, Y. Dziashko, R. Ding, S. Fathololoumi, A. E.-J. Lim, K. Padmaraju, R. Sukkar, R. Younce, H. Rohde, R. Palmer, G. Saathoff, T. Wuth, M. Bohn, A. Ahmed, M. Ahmed, C. Williams, D. Lim, A. Elmoznine, A. Rylyakov, T. Baehr-Jones, P. Magill, D. Scordo, and M. Hochberg, "A silicon photonic transceiver and hybrid tunable laser for 64 Gbaud coherent communication," in *Optical Fiber Communication Conference Postdeadline Papers* (OSA, 2018), paper Th4D.4.
20. C. Kopp, S. Bernabé, B. Ben Bakir, J. Fedeli, R. Orobtchouk, F. Schrank, H. Porte, L. Zimmermann, and T. Tekin, "Silicon photonic circuits: on-CMOS integration, fiber optical coupling, and packaging," IEEE J. Sel. Top. Quantum Electron. **17**(3), 498–509 (2011).
21. E. Higurashi, T. Imamura, T. Suga, and R. Sawada, "Low-temperature bonding of laser diode chips on silicon substrates using plasma activation of Au films," IEEE Photonics Technol. Lett. **19**(24), 1994–1996 (2007).

22. S. Bernabe, B. Blampey, A. Myko, B. Charbonnier, K. Frigui, S. Bila, A. Mottet, J. Hauden, S. Jillard, B. Frigui, G. Duan, X. Pommarede, and G. Charlet, "Packaging of photonic integrated circuit based high-speed coherent transmitter module," in *2016 IEEE 66th Electronic Components and Technology Conference (ECTC)* (IEEE, 2016), pp. 1081–1086.
23. S. Bernabé, S. Olivier, A. Myko, M. Fournier, B. Blampey, A. Abraham, S. Menezo, J. Hauden, A. Mottet, K. Frigui, S. Ngoho, B. Frigui, S. Bila, D. Marris-Morini, D. Pérez-Galacho, P. Brindel, and G. Charlet, "High-speed coherent silicon modulator module using photonic integrated circuits: from circuit design to packaged module," in *Proceedings of SPIE - The International Society for Optical Engineering* (SPIE Photonics Europe, 2016), paoper 98910Z.
24. D. Celo, D. J. Goodwill, J. Jiang, P. Dumais, C. Zhang, F. Zhao, X. Tu, C. Zhang, S. Yan, J. He, M. Li, W. Liu, Y. Wei, D. Geng, H. Mehrvar, and E. Bernier, "32x32 silicon photonic switch," in *OptoElectronics and Communications Conference (OECC) Held Jointly with 2016 International Conference on Photonics in Switching (PS)* (2016), paper WF1–4.
25. C. Hoessbacher, Y. Salamin, Y. Fedoryshyn, W. Heni, B. Baeuerle, A. Josten, C. Haffner, M. Zahner, H. Chen, D. L. Elder, S. Wehrli, D. Hillerkuss, D. Van Thourhout, J. Van Campenhout, L. R. Dalton, C. Hafner, and J. Leuthold, "Optical interconnect solution with plasmonic modulator and Ge photodetector array," IEEE Photonics Technol. Lett. **29**(21), 1760–1763 (2017).
26. D. Vermeulen, R. Aroca, L. Chen, L. Pellach, G. McBrien, and C. Doerr, "Demonstration of silicon photonics push–pull modulators designed for manufacturability," IEEE Photonics Technol. Lett. **28**(10), 1127–1129 (2016).
27. C. Lacava, I. Cardea, I. Demirtzioglou, A. E. Khoja, L. Ke, D. J. Thomson, X. Ruan, F. Zhang, G. T. Reed, D. J. Richardson, and P. Petropoulos, "49.6 Gb/s direct detection DMT transmission over 40 km single mode fibre using an electrically packaged silicon photonic modulator," Opt. Express **25**(24), 29798–29811 (2017).
28. X. Ruan, K. Li, D. J. Thomson, C. Lacava, F. Meng, I. Demirtzioglou, P. Petropoulos, Y. Zhu, G. T. Reed, and F. Zhang, "Experimental comparison of direct detection Nyquist SSB transmission based on silicon dual-drive and IQ Mach-Zehnder modulators with electrical packaging," Opt. Express **25**(16), 19332–19342 (2017).
29. P. Dong, X. Liu, S. Chandrasekhar, L. L. Buhl, R. Aroca, and Y.-K. Chen, "Monolithic silicon photonic integrated circuits for compact 100+Gb/s coherent optical receivers and transmitters," IEEE J. Sel. Top. Quantum Electron. **20**, 150–157 (2014).
30. N. Qi, X. Xiao, S. Hu, X. Li, H. Li, L. Liu, Z. Li, N. Wu, and P. Y. Chiang, "Co-Design and demonstration of a 25-Gb/s silicon-photonic Mach–Zehnder modulator with a CMOS-based high-swing driver," IEEE J. Sel. Top. Quantum Electron. **22**(6), 131–140 (2016).
31. Y. Ding, C. Peucheret, H. Ou, and K. Yvind, "Fully etched apodized grating coupler on the SOI platform with -0.58 dB coupling efficiency," Opt. Lett. **39**(18), 5348–5350 (2014).
32. N. Lindenmann, S. Dottermusch, M. L. Goedecke, T. Hoose, M. R. Billah, T. P. Onanuga, A. Hofmann, W. Freude, and C. Koos, "Connecting silicon photonic circuits to multicore fibers by photonic wire bonding," J. Lightwave Technol. **33**(4), 755–760 (2015).
33. P.-I. Dietrich, M. Blaicher, I. Reuter, M. Billah, T. Hoose, A. Hofmann, C. Caer, R. Dangel, B. Offrein, U. Troppenz, M. Moehrle, W. Freude, and C. Koos, "In situ 3D nanoprinting of free-form coupling elements for hybrid photonic integration," Nat. Photonics **12**(4), 241–247 (2018).
34. H. Zwickel, J. N. Kemal, C. Kieninger, Y. Kutuvantavida, M. Lauermann, J. Rittershofer, R. Pajković, D. Lindt, S. Randel, W. Freude, and C. Koos, "Electrically Packaged Silicon-Organic Hybrid Modulator for Communication and Microwave Photonic Applications," in *Conference on Lasers and Electro-Optics* (OSA, 2018), paper SM3B.1.
35. H. Miura, F. Qiu, A. M. Spring, T. Kashino, T. Kikuchi, M. Ozawa, H. Nawata, K. Odoi, and S. Yokoyama, "High thermal stability 40 GHz electro-optic polymer modulators," Opt. Express **25**(23), 28643–28649 (2017).
36. C. Kieninger, Y. Kutuvantavida, J. N. Kemal, H. Zwickel, H. Miura, S. Randel, W. Freude, S. Yokoyama, and C. Koos, "Demonstration of long-term thermal stability of a silicon-organic hybrid (SOH) modulator at 85°C," Opt. Express **26**, 27955–27964 (2018).
37. M. Billah, J. N. Kemal, M. Blaicher, Y. Kutuvantavida, C. Kieninger, H. Zwickel, P.-I. Dietrich, S. Wolf, T. Hoose, Y. Xu, U. Troppenz, M. Möhrle, S. Randel, W. Freude, and C. Koos, "Four-channel 784 Gbit / s transmitter module enabled by photonic wire bonding and silicon-organic hybrid modulators," in *European Conference on Optical Communication (ECOC)* (2017), paper Th.PDP.C.1.
38. R. Ding, T. Baehr-Jones, Y. Liu, R. Bojko, J. Witzens, S. Huang, J. Luo, S. Benight, P. Sullivan, J.-M. Fedeli, M. Fournier, L. Dalton, A. Jen, and M. Hochberg, "Demonstration of a low V π L modulator with GHz bandwidth based on electro-optic polymer-clad silicon slot waveguides," Opt. Express **18**(15), 15618–15623 (2010).
39. C. Koos, J. Brosi, M. Waldow, W. Freude, and J. Leuthold, "Silicon-on-insulator modulators for next-generation 100 Gbit/s-ethernet," in *European Conference on Optical Communication (ECOC)* (2007), paper P056.
40. M. J. Degerstrom, B. K. Gilbert, and E. S. Daniel, "Accurate resistance, inductance, capacitance, and conductance (RLCG) from uniform transmission line measurements," in *2008 IEEE-EPEP Electrical Performance of Electronic Packaging* (IEEE, 2008), pp. 77–80.
41. F. J. Schmückle, R. Doerner, G. N. Phung, W. Heinrich, D. Williams, and U. Arz, "Radiation, Multimode Propagation, and Substrate Modes in W-Band CPW Calibrations," in *41st European Microwave Conference* (2011), pp. 297–300.

42. J. Fujikata, M. Noguchi, J. Han, S. Takahashi, M. Takenaka, and T. Nakamura, "Record-high modulation-efficiency depletion-type Si-based optical modulator with in-situ B doped strained SiGe layer on Si waveguide for 1.3 um wavelength," in *European Conference on Optical Communications (ECOC)* (2016), paper Tu.3.A.4.
43. L. Alloatti, R. Palmer, S. Diebold, K. P. Pahl, B. Chen, R. Dinu, M. Fournier, J.-M. Fedeli, T. Zwick, W. Freude, C. Koos, and J. Leuthold, "100 GHz silicon–organic hybrid modulator," Light Sci. Appl. **3**(5), e173 (2014).
44. L. Alloatti, D. Korn, R. Palmer, D. Hillerkuss, J. Li, A. Barklund, R. Dinu, J. Wieland, M. Fournier, J. Fedeli, H. Yu, W. Bogaerts, P. Dumon, R. Baets, C. Koos, W. Freude, and J. Leuthold, "42.7 Gbit/s electro-optic modulator in silicon technology," Opt. Express **19**(12), 11841–11851 (2011).
45. S. Wolf, H. Zwickel, W. Hartmann, M. Lauermann, Y. Kutuvantavida, C. Kieninger, L. Altenhain, R. Schmid, J. Luo, A. K.-Y. Jen, S. Randel, W. Freude, and C. Koos, "Silicon-organic hybrid (SOH) Mach-Zehnder modulators for 100 Gbit/s on-off keying," Sci. Rep. **8**(1), 2598 (2018).
46. W. Freude, R. Schmogrow, B. Nebendahl, M. Winter, A. Josten, D. Hillerkuss, S. Koenig, J. Meyer, M. Dreschmann, M. Huebner, C. Koos, J. Becker, and J. Leuthold, "Quality metrics for optical signals: Eye diagram, Q-factor, OSNR, EVM and BER," in *International Conference on Transparent Optical Networks (ICTON)* (IEEE, 2012), paper Mo.B1.5.
47. L. M. Zhang and F. R. Kschischang, "Staircase codes with 6% to 33% overhead," J. Lightwave Technol. **32**(10), 1999–2002 (2014).